\documentclass[twocolumn,article,superscriptaddress,10pt]{revtex4-2}
\usepackage{amsmath}
\usepackage{amssymb}
\usepackage{graphicx}
\usepackage{xcolor}

\begin{document}

\title{Two-dimensional non-linear hydrodynamics and nanofluidics}

\author{Maxim Trushin}
\email{mxt@nus.edu.sg}
\affiliation{Institute for Functional Intelligent Materials, National University of Singapore, Singapore 117544}
\affiliation{Centre for Advanced 2D Materials, National University of Singapore, Singapore 117546}
\affiliation{Department of Materials Science and Engineering, National University of Singapore, Singapore 117575 }

\author{Alexandra Carvalho}
\email{carvalho@nus.edu.sg}
\affiliation{Institute for Functional Intelligent Materials, National University of Singapore, Singapore 117544}
\affiliation{Centre for Advanced 2D Materials, National University of Singapore, Singapore 117546}

\author{A. H. Castro Neto}
\affiliation{Institute for Functional Intelligent Materials, National University of Singapore, Singapore 117544}
\affiliation{Centre for Advanced 2D Materials, National University of Singapore, Singapore 117546}
\affiliation{Department of Materials Science and Engineering, National University of Singapore, Singapore 117575 }

\begin{abstract}
{\bf Abstract}--- A water monolayer squeezed between two solid planes 
experiences strong out-of-plane confinement effects while expanding freely within the plane.
As a consequence, the transport of such two-dimensional water combines hydrodynamic and nanofluidic features, intimately linked with each other.
In this paper, we propose and explicitly solve a non-linear hydrodynamic equation describing two-dimensional water flow
with viscosity parameters deduced from molecular dynamic simulations. We demonstrate
that the very ability of two-dimensional water 
to flow in short channels is governed by the second (dilatational) viscosity coefficient, leading to flow compression and velocity saturation
in the high-pressure limit.
The viscosity parameter values depend strongly on whether graphene or hexoganal boron nitride layers are used to confine 2D water that offers
an interesting opportunity to obtain various nanofluids out of the same water molecules just by using alternate materials to fabricate the 2D channels.

\end{abstract}

\maketitle

\section*{Introduction}

Water is the most important substance for life on Earth and has remained in the scientific focus 
for centuries, if not for millennia.
Despite (or due to) its simple chemical composition, the structure of water often causes
scientific controversies such as polywater \cite{derjaguin1983polywater},
memory effect \cite{maddox1988high},  chain-like formation of water molecules \cite{head2006tetrahedral}, to mention a few. 
The most controversial claims regarding non-orthodoxal properties of bulk water have been debunked by subsequent comprehensive experiments \cite{derjaguin1983polywater,maddox1988high,head2006tetrahedral}.
It may nevertheless be possible to find some evidence of unconventional phenomena in 
two-dimensional (2D) water --- a-few-angstrom thick water monolayer squeezed between two solid planes \cite{Geim2019}.

What makes water so special as compared with most other liquids are the relatively strong hydrogen bonds \cite{smith2005unified}. 
In particular, the hydrogen bonds are believed to be responsible for the rather high melting and boiling points of water,
as well as for the expansion upon freezing.
In general, reducing dimensionality of any interacting physical system (for instance, squeezing it into a plane) 
amplifies interaction effects. 
The same happens in 2D water: The hydrogen bonds may become stable enough to bring
water molecules into an ordered state making water behave like a solid in some aspects.
2D water locked between two graphene sheets has been recently found in such a structured state \cite{algara2015square}.
Molecular dynamics (MD) \cite{gao2018phase,qiu2015water,mario2015aa,fernandez2016electric,PCCP2017diao,corsetti2016enhanced} and ab-initio simulations \cite{zangi2003monolayer,corsetti2016enhanced,chen2016two,corsetti2016structural,kapil2022first,ghorbanfekr2020insights} suggest that
2D water can transit into various structured states with distinctive molecular arrangements\cite{zhao2014highly}.
Recent progress in machine learning has made it possible to overcome some limitations of ab-initio and force field methods \cite{MachLearn1,MachLearn2,MachLearn3,kapil2022first}.
Nonetheless, hydrodynamics of truly 2D water remains an open question \cite{kavokine2021fluids}.

Until recently, the fundamental research on strongly confined water has mostly focused on carbon nanotubes \cite{majumder2005enhanced,holt2006fast,NatureNano2017}
with a diameter of less than 1 nm
--- a characteristic length scale below which any continuum (hydrodynamic) description generally fails \cite{bocquet2010nanofluidics}.
At such a small scale the finite-size effects associated with molecule geometry and channel diameter become crucial 
and must be probed using nanofluidic methods, such as MD simulations \cite{hummer2001water,mukherjee2010single,su2011control}.
In 2D water, the in-plane flow remains essentially unrestricted, hence,
it should follow the laws of hydrodynamics, in which the finite size of molecules never enters explicitly.
At the same time, the out-of-plane molecular motion is strongly restricted and
falls into the realm of nanofluidics. 2D water can therefore be seen as a hybrid system having hydrodynamic features in the in-plane directions and nanofluidic 
features along the out-of-plane direction, which can affect each other in some ways not known so far.

An attempt to understand such a hybridization has been recently made using a Poiseuille-like model with 
the viscosity coefficients taken from MD simulations \cite{APL2018peeters}.
Although a certain consistency between the continuum model and MD simulations has been reached, the very applicability of the Poiseuille equation to a monolayer remains questionable.
Indeed, what is usually considered as a confined 2D water flow is in fact a quasi 2D one, where no water monolayer is formed, Fig. \ref{figure1}a.
The flow is assumed to be laminar, and the outer layers being in contact with the walls travel at a slower velocity than the inner layers,
resulting in an out-of-plane velocity profile, $v_x(z)$, absent in the truly 2D limit. 
The profile determines the slip length, $l_\mathrm{s}$, which could also be defined as a ratio 
between the bulk shear viscosity and interfacial friction coefficients
\cite{PRE2014sliplength,petravic2007equilibrium,JoCP2021} so that
it characterizes the relative contributions of the bulk and interface frictions into energy dissipation \cite{bocquet2007flow}.
Obviously, it is not possible to distinguish between the bulk and interface in a 2D limit, and
the shear viscosity and interfacial friction coefficient either lose any sense or must be redefined.
Hence, a proper continuum model for 2D water cannot be directly deduced from the conventional models used so far.

We offer an alternative to the Poiseuille formula, also relating driving pressure,
flow velocity, and viscosity coefficients, but suitable for truly 2D water (Fig. \ref{figure1}b).
It is given by an explicit solution of the non-linear hydrodynamic equation written as
\begin{equation}
 \label{general}
 \eta \frac{\partial^2 v}{\partial x^2} + v_0\rho_0 \left(\frac{c_0^2}{v^2} - 1\right) \frac{\partial v}{\partial x}=0,
\end{equation}
where $v$ is the flow velocity along the coordinate $x$, $\eta=4\eta_1/3 +\eta_2$ is the total viscosity
with $\eta_{1,2}$ being the viscosity coefficients discussed below,
$c_0$ is the sound velocity, $v_0$ and $\rho_0$ are the flow velocity and density at $x=0$, respectively. 
Equation (\ref{general}) is derived in this paper from the 2D Navier-Stokes and continuity equations
assuming a well-structured flow (i.e. no vorticity, $\mathrm{rot}\, \mathbf{v} =0$)
with a certain compressibility ($\mathrm{div}\, \mathbf{v}\neq 0$) in short channels, where conventional hydrodynamic friction effects can be neglected.

In Eq. (\ref{general}), $\eta_1$ does not represent conventional (intrinsic) shear viscosity,
as the absence of vorticity implies that 2D water is a solid rather than a liquid
so that the intrinsic shear viscosity coefficient would formally be infinite 
(or demonstrate tens of orders of magnitude increase, as compared to a liquid state \cite{glass_viscosity_1996}), and
it is known that conventional shear viscosity of water increases by orders of magnitude
when approaching a monolayer limit \cite{neek2016commensurability}.
In our model, $\eta_1$ has the meaning of interfacial viscosity, i.e. it is determined by interactions
between water and the solid layers forming the 2D channel.
The second viscosity coefficient $\eta_{2}$ is associated with the energy loss caused 
by compression or expansion of the water monolayer and referred here to as dilatational viscosity, $\eta_2$.
In contrast to the conventional bulk viscosity
\cite{liebermann1949second,dukhin2009bulk,jaeger2018bulk}, the dilatational viscosity is of utmost importance for 2D water flow
because the water molecules confined in 2D tend to have a denser hydrogen bonding network 
under stronger interaction with the solid surface \cite{sendner2009interfacial,qin2015nonlinear,maekawa2018structure}, and
the resulting density may also vary under the stress \cite{water-slip-stress-2016}
and structural changes \cite{neek2016commensurability,PCCP2017diao}.
Hence, neither of $\eta_{1,2}$ is intrinsic, and their values depend on the material the channel is made of.
Both coefficients appear as a linear combination in Eq. (\ref{general}) but they enter the boundary 
conditions separately.

In this work, Eq. (\ref{general}) is explicitly solved in the limit $v(x)\ll c_0$ under the boundary conditions at $x=0$ for $v(x)$ and $\partial_x v(x)$
determined by a driving pressure. Using MD simulations, we show that our solution represents a realistic model for the 2D water flow 
confined by carbon or boron nitride planes in the channels of up to 10 nm length.
We also show that the interfacial and dilatational viscosity parameters $\eta_1$ and $\eta_2$ 
are not intrinsic to 2D water but strongly affected by the material the 2D channel is made of.
The interplay between hydrodynamic and nanofluidic mechanisms
leading to the non-linear dependence of the 2D water flow velocity on driving pressure and the viscosity parameters is the main focus of this research.

\section*{Results}

\subsection*{Model setup}
\label{model}

\begin{figure}
 \includegraphics[width=\columnwidth]{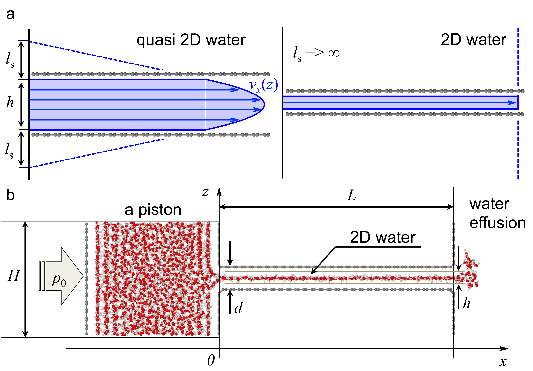}
 \caption{{\bf Hydrodynamic and molecular dynamic (MD) setups for two-dimensional (2D) water. }
 {\bf a} Difference between quasi 2D water, described in terms of the slip length $l_\mathrm{s}$, and 2D water studied here:
 The former has a well-defined out-of-plane velocity profile $v_x(z)$, whereas the latter does not.
 Formally speaking, the slip length is infinite for 2D water and therefore unsuitable for the characterisation of a truly 2D flow.
  {\bf b} Our MD setup for testing solution of Eq. (\ref{general}):
  A microscopic piston of height $H$ pushes water through a 2D channel of length $L$ at a constant pressure $p_0$.
Here, the structure is made of carbon but we also consider hexagonal boron nitride as an option.
 The channel height (carbon-carbon interlayer distance, $d$) is chosen to be small enough (6 or 7 \AA) to squeeze water down to a monolayer.
 The hydrodynamic channel height $h = d - 2\delta_\mathrm{vdW}$ takes into account the van der Waals off-set $\delta_\mathrm{vdW}=1.67\, \mathrm{\mathring{A}}$ from each side
 of the channel not accessible for water molecules.
 The piston cross-section is assumed to be substantially larger than that of the channel ($H \sim 30\, \mathrm{\mathring{A}}$).
 The exit of the channel is assumed to be always opened so that the boundary conditions are imposed at $x=0$.
 An animated version of our setup is available as Supplementary Movie 1. 
 }
 \label{figure1}
\end{figure}

 To test our equation (\ref{general}) by means of MD simulations we consider a channel of height $h$ formed by two parallel solid planes of length $L$. The 2D water flowing in this region will be the object of this study.
However, in practice the 2D water flow has to be fed by a source. In the molecular dynamics simulations, the channel connects two reservoirs: one is always full of water, and another is almost empty.
A piston of height $H$ is moving to maintain a constant driving pressure $p_0$.
The piston and channel have the same width $w$ ($w\gg h,H$), with cross-sections are $wH$ and $wh$, respectively.
The problem is effectively one-dimensional with the coordinate $x$ directed along the flow and origin $x=0$ defined to be at the  entrance of the channel, see Fig. \ref{figure1}b.
Note that the hydrodynamic height $h$ is smaller than the actual MD interlayer distance $d$ measured from the middle of the upper solid layer to the middle of the lower one, due to the fact that the electronic orbitals on both sides narrow the hydrodynamic channel. We estimate this narrowing to be by $1.67\, \mathrm{\mathring{A}}$ from each side (half the graphite or h-BN interlayer distance).

A steady-state fluid flow velocity is described by a second-order differential equation, the Navier-Stokes equation (see Methods), 
and its solution requires two boundary conditions.
Physically, the boundary conditions take into account the feeding reservoir,
which is out of scope of our 2D hydrodynamic theory, but must be retained in our MD simulations.
Hence, the choice of $v_0$ is specific to the particular MD simulation setup we are currently utilizing.
Testing a few reasonable relations between driving pressure and $v_0$ we 
have found the best fit is given by the simplest Bernoulli's equation as $p_0=\rho v^2/2|_{x=0^-}$
assuming $\rho_{x=0^-}=\rho_{x=0}=\rho_0$.
The flow velocity is then higher at the entrance point ($x=0$) than in the left reservoir ($x<0$) because of the mass conservation
equation given by $wH \rho_0 v|_{x=0^-} = w h \rho_0 v|_{x=0}$. Hence, we have
\begin{equation}
\left. v\right|_{x=0} \equiv v_0 = \frac{H}{h}\sqrt{\frac{2 p_0}{\rho_0}}.
\label{1st}
\end{equation}
Equation (\ref{1st}) could be modified by means of the Darcy--Weisbach relation \cite{darcy-weisbach} with a phenomenological friction factor.
Technically, the friction factor could be absorbed into the effective hydrodynamic channel height $h$.
Note, that the Darcy--Weisbach equation also suggests quadratic relation between pressure and average flow velocity, so that
the functional dependence would be the same as in Eq. (2). 
We shall see later that the average 2D flow velocity obtained from our MD simulations indeed tends to follow a square-root dependence 
on the driving pressure as soon as the channel height becomes larger than the hydrodynamic limit of about 1 nm.
In that way, applicability of Eq. (\ref{1st}) is justified, and its simplicity 
can be explained by the peculiarities of our MD setup, where the piston has no walls along the flow, 
as we apply periodic boundary conditions in $y$ and $z$ directions at $x<0$. 
Hence, there is no energy loss associated with the walls. The energy loss due to water compression at the entrance ($x=0$)
is taken into account by the second boundary condition, as follows.

The second boundary condition applies to the divergence of the flow velocity, $\mathrm{div}\, \mathbf{v}$.
If $\mathrm{div}\, \mathbf{v}\neq 0$, then the continuity equation immediately suggests that $\mathrm{grad}\,\rho \neq 0$
requiring 2D water to be able to shrink and expand. 
It is the second (dilatational)  viscosity \cite{landau-hydrodynamics} that relates the pressure difference and $\mathrm{div}\, \mathbf{v}$ 
in a steady-state limit as $p_0 = -\eta_2 \mathrm{div}\, \mathbf{v}$.
Our $p_0$ is not to confuse with the equilibrium pressure also denoted by $p_0$ in \cite{landau-hydrodynamics}.
The latter is nearly zero in our case because the right volume in Fig. \ref{figure1}b, is very large and almost empty.
For our effectively one-dimensional problem the second boundary condition can be written as
\begin{equation}
 \left. \frac{\partial v}{\partial x} \right|_{x=0} = -\frac{p_0}{\eta_2}.
 \label{2nd}
\end{equation}
This expression relates the compressibility of the fluid to an external 
perturbation (driving pressure) via a material parameter (dilatational viscosity)
and in that way describes viscous entrance effects.
Since $\eta_2>0$ and $p_0>0$ the flow must slow down when propagating through at least the starting section of the channel.
Hence, the flow density must increase with $x$.
The limit of $\eta_2\to\infty$ corresponds to an ideal solid state,
when the ice layer never deforms regardless of the stress applied. The flow velocity then does not change within the channel limits. The opposite limit of $\eta_2\to 0$ corresponds to an ideal gas state with an absolute compressibility resulting in a vanishing flow velocity right at the entrance of the channel.
Eq.~(\ref{2nd}) also suggests that such a low dilatational viscosity coefficient leads to a high responsivity to driving pressure anticipating non-linear effects. The 2D water flow characteristics are supposed to lie between these two limits depending on the interactions within the channel.
We emphasize that the non-linear effects are intrinsic to our model regardless of the boundary conditions.

Right after entering the 2D channel the water flow becomes strongly confined in $z$-direction with both density and velocity being dependent on $x$.
Indeed, the statistical analysis of our MD simulation data (Supplementary Figures 1--4) indicates that
(i) the averaged $\langle v_x \rangle$ is substantially higher than $\langle v_z\rangle$; 
(ii) the averaged exit velocity $\langle v_{x=L} \rangle$ is somewhat lower than the averaged entry velocity, $\langle v_{x=0} \rangle$;
(iii) increasing $d$ to 1 nm makes the averaged flow velocity $\langle v(x) \rangle_L$ equal to $v_0$ indicating transition to the conventional regime at $d>1$ nm;
(iv) the flow velocity out-of-plane profile is rectangular rather than parabolic prohibiting description in terms of the slip length $l_\mathrm{s}$.
Note that the left reservoir having $H > 1$ nm is always in the conventional (Bernoulli's) regime, except when the piston approaches $x=0$.

\begin{figure*}
\includegraphics[width=\textwidth]{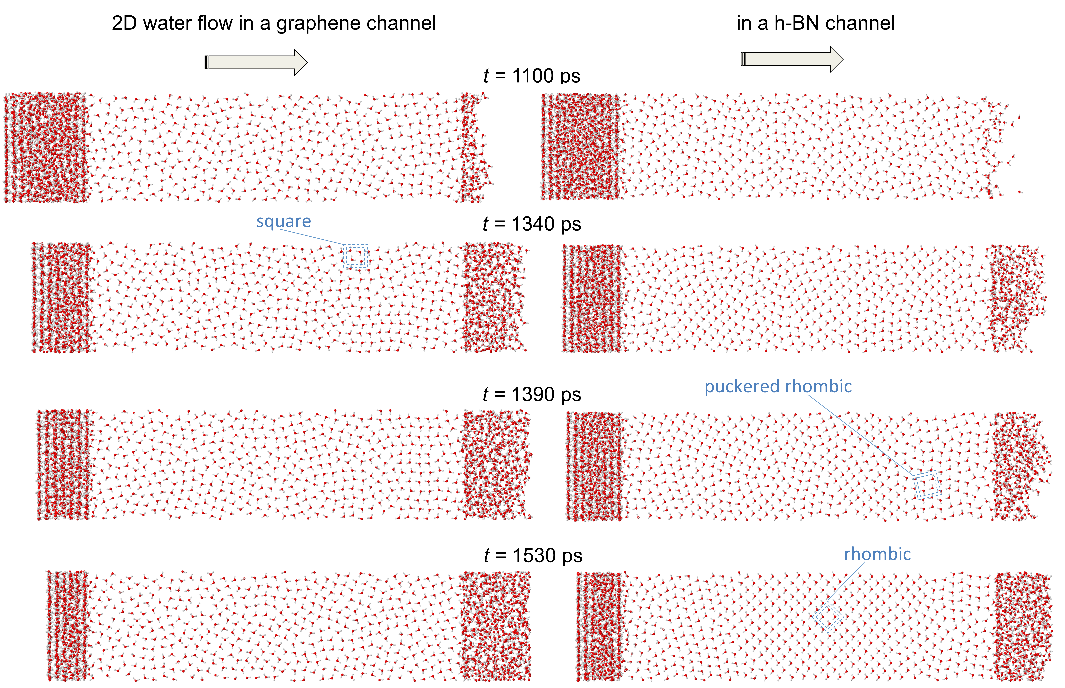}
 \caption{{\bf Molecular dynamic (MD) top-view snapshots showing different two-dimensional (2D) water structures developing along graphene 
 and hexagonal boron nitride (h-BN) channels.}
 All water molecules are placed into the left reservoir at $t=0$. The snapshots are made at the indicated time points.
 The typical patterns are highlighted.
 Note that the crystalline regions move with the flow, and different grain patterns appear at different time steps.
 The channel height is $d=6\,\mathrm{\mathring{A}}$, and the atoms other than H (white) and O (red) are removed for clarity.
 The water molecules appear to be more ordered in h-BN than in graphene channels. An animated version of 2D water flow in a graphene channel is available as Supplementary Movie 2.}
 \label{figure2}
\end{figure*}

To relate the changes of pressure and density we introduce the speed of sound defined as
$c_0= \sqrt{{\partial p}/{\partial \rho}}$.
Here, $c_0$ is assumed to be a constant with the conventional value $1.5\cdot 10^3$ m/s.
In the channel, the speed of sound and viscosity coefficients may depend on $x$,
 but we assume constant values in order to keep the hydrodynamic model analytically tractable.
As we shall see below, the qualitative outcomes are not sensitive to this assumption.

\subsection*{The structure of 2D water}
\label{structure}

In this section, we consider the structure adopted by water when confined in the channel, in the stationary flow regime.
Figure \ref{figure2} suggests that 2D water rapidly develops a multicrystalline structure when entering the channel.
A solid phase of 2D water is expected at room temperature, as it has been shown that for water confined by graphene, there is a solid-liquid phase transition at an interlayer distance of about 7.5~\AA, for a density of about 12 molecules/nm$^2$ such as the one that we have used here \cite{gao2018phase}. We have observed a similar phase transition for stationary 2D water MD models, confirming that this is not a problem of the thermostating such as the `flying-ice-cube' problem discussed previously \cite{harvey-ice-cube}.
We have observed regions of the square phase and rhombic polar phase for graphene confinement, and rhombic polar and square polar for boron nitride confinement (for nomenclature please refer to Rf.~\cite{negi2022edge}), consistent with previous works \cite{chen2016two,zangi2003monolayer,gao2018phase,li2019two,corsetti2016enhanced,corsetti2016structural,kapil2022first,ghorbanfekr2020insights}, in agreement with the experimental observation of `square ice' for monolayer water confined by graphene at room temperature \cite{algara2015square}.
The hydrogen atoms are less ordered than the oxygen atoms, as typical of other ice phases, in 3D.

\begin{figure}
 \includegraphics[width=\columnwidth]{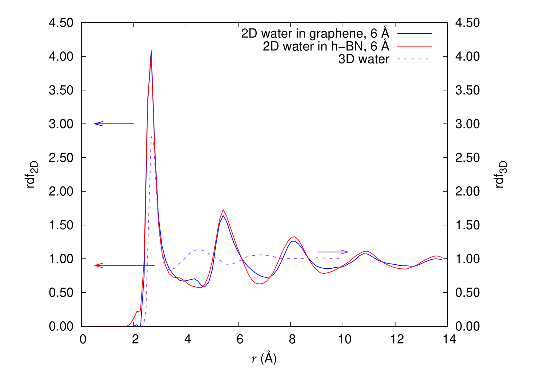}
 \caption{{\bf The O-O radial distribution function (rdf) for two-dimensional (2D) water.} The water layer is confined between graphene or h-BN planes
 at a pressure of 5.2 atm. The O-O radial distribution function for static bulk liquid water is given as well, for comparison.
The rdf$_\mathrm{2D}$ is computed for the water confined to the area of the 2D channel, averaged over time, see ``Molecular dynamics simulations'' in Methods. 
The rdf$_\mathrm{3D}$ for 3D water is calculated using the regular 3D expression, for a cube of water of 35.6~\AA~side.}
 \label{figure3}
\end{figure}

The most striking difference between carbon and BN channels is the domain size of the 2D ice crystal regions.
The domains tend to be larger in BN channels, hence, 2D water appears to be more structured by BN walls than by carbon ones.
To quantify the crystallinity and long-range order we have calculated the radial distribution function (rdf, see Methods).
For bulk liquid water, the (3D) radial distribution function shows a sharp peak at 2.8~\AA, corresponding to the nearest-neighbour distance, and decays fast, 
showing still two more peaks at approximately 4.4 and 6.7 \AA, see Fig. \ref{figure3}.
In contrast, the rdf for 2D water shows multiple peaks as a function of $r$ in both h-BN and carbon channels clearly indicating long-range order. 
The peak corresponding to the nearest-neighbour shell coincides for the cases of graphene and h-BN channels and is located at 2.6~\AA.
In the case of graphene, there is a smaller feature at about 4.4~\AA. 
This is too low to correspond  to the second nearest neighbour shell for a perfect square lattice, 
but it is very close to the value expected for a 60\textdegree-rhombus, $\sqrt{3} \times 2.6$\AA.
The other peaks are comparatively more pronounced in the case of h-BN channels,
as expected for the more structured molecular arrangements (compare with graphene in Fig. \ref{figure2}). 
Hence, graphene and h-BN interact with water differently, which should result
in different interfacial and dilatational viscosity coefficients.

\subsection*{2D hydrodynamics}

We expect 2D water and bulk liquid water to flow differently.
Our MD simulations suggest that 2D water maintains its ordered state
when moving in the channel, and interfacial viscosity due to interactions with the channel's walls
can be stronger than interactions between adjacent layers in the laminar flow of bulk water.
Hence, the interfacial viscosity coefficient, $\eta_1$, is expected to be somewhat larger than the bulk water viscosity of about $1$ mPa$\cdot$s \cite{jaeger2018bulk}.
In contrast, the dilatational viscosity coefficient, $\eta_2$, should be much lower
because of the stronger compressibility of 2D water.
The in-plane compressibility and the absence of laminar structure can be formalised by means of the Navier-Stokes equation, see Methods.
To our best knowledge, the resulting Eq. (\ref{general}) has never occurred in the literature despite 
having some distant formal similarities with Chaplygin's equation \cite{landau-hydrodynamics} describing 
a steady-state potential flow of a 2D compressible gas.
In Methods, we show that the solution of Eq. (\ref{general}) can be parameterised in terms of $\gamma$ and $\tau$ given by
\begin{equation}
 \gamma =\frac{\frac{\eta_2}{\eta}\rho_0 c_0^2}{p_0 + \frac{\eta_2}{\eta}\rho_0 c_0^2},
 \label{gamma}
\end{equation}
and
\begin{equation}
 \tau = \frac{\eta_2}{p_0 + \frac{\eta_2}{\eta}\rho_0 c_0^2}.
 \label{tau}
\end{equation}
It is the parameter $\gamma$ that is responsible for the non-linear effects. 
If $\gamma=1$, then $v=v_0$, and the water layer neither shrinks nor expands when propagating through the channel,
which corresponds to an infinitesimally small driving pressure or 
infinitely high dilatational viscosity.
If $\gamma > 1$, then the water flow velocity increases with $x$.
Equation (\ref{gamma}) suggests that such a regime is obviously impossible.
If $\gamma < 1$, then the water flow slows down in the channel.
Having in mind that $\rho_0 c_0^2\sim 2.2\cdot10^4$ atm the realistic values of $\gamma$ are just slightly below $1$ at any reasonable driving pressure and viscosity coefficients.

It is easy to understand the physical meaning of $\eta_{1,2}$ by considering the limiting cases.
If $\eta_1 \to \infty$ but $\eta_2$ remains finite, then the water layer cannot slide, and the flow is stuck ($\gamma\to 0$).
One can imagine that the interfacial viscosity is so high as if the water layer and the channel's walls are glued together.
If $\eta_2 \to 0$ but $\eta_1$ remains finite, then the water layer cannot resist compression, and the flow is stuck again ($\gamma\to 0$).
The density would then formally diverge at the entrance of the channel. Physically, the flow would be jammed.
In this limiting case, 2D water layer behaves like a soft rubber band pushed through a narrow channel: it obviously crumples and cannot get through.
The coefficients $\eta_{1,2}$ describe the two mechanisms potentially limiting 2D water transport.

\begin{figure}
\includegraphics[width=\columnwidth]{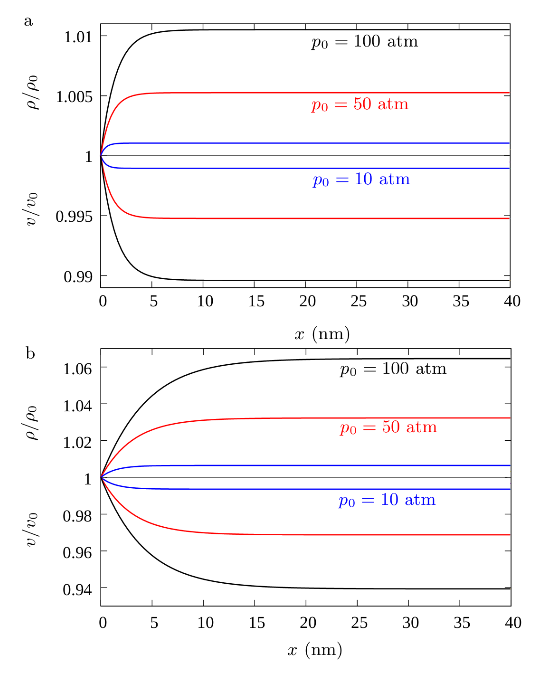}
 \caption{{\bf Density and velocity of two-dimensional (2D) water flow within our hydrodynamic model.} 
 The flow velocity saturates when moving along the channel as described by Eq. (\ref{general})
 with its solution given in terms of the Lambert functions, see ``Non-linear hydrodynamics in a 2D channel'' in Methods.
 The saturated values, $v_0\gamma$ and $\rho_0/\gamma$, depend on the viscosity parameters.
 {\bf a} The interfacial and dilatational viscosity coefficients are equal: $\eta_{1,2} \sim 1$ mPa$\cdot$s.
 {\bf b} The viscosity coefficients are unequal: $\eta_1 = 5$ mPa$\cdot$s, $\eta_2 = 0.5$ mPa$\cdot$s.
 The remaining parameters are $\rho_0=1\cdot 10^3$ kg/m$^3$, $c_0=1.5\cdot 10^3$ m/s.
 Note that a certain change of $\rho_0 c_0^2$  can be compensated by adjusting the ratio $\eta_2/\eta$ to obtain the same value for $\gamma$, if necessary.
 The driving pressure is shown in the figure.}
 \label{figure4}
\end{figure}

Figure \ref{figure4} shows that $v(x)$ drops down with increasing $x$, and $\rho(x)$ steps up accordingly.
The local pressure also increases with $x$ but
the global pressure difference between the left and right reservoirs remains positive providing continuous flow. 
The velocity maximum is at $x=0$, as $v(0) = v_0$, and the velocity minimum can be seen as 
$v(\infty) = v_0\gamma$. The higher driving pressure results in the larger difference $v(0)-v(\infty)$.
Figure \ref{figure4}a suggests that the water density $\rho(x)$ changes just by about 1\% at $\eta_1 \sim \eta_2$
within the channel length even though driving pressure up to 100 atm is applied.
The compression is facilitated when $\eta_1$ increases and $\eta_2$ decreases, see Fig. \ref{figure4}b.

Another parameter to discuss is the length $v_0\tau$ with $\tau$ given by Eq. (\ref{tau}).
This is the characteristic distance, measured from the channel's entrance, within which 
 both flow velocity and density are approximately saturated at their respective values, $v_0\gamma$ and $\rho_0/\gamma$.
The length is determined by the driving pressure as well as by dilatational viscosity.
If $\eta_{1,2}\sim 1$ mPa$\cdot$s, then $\tau\sim 1$ ps at low pressure,
and assuming $v$ of a few $\mathrm{\mathring{A}/ps}$ we obtain the characteristic length of 
a few $\mathrm{\mathring{A}}$. The length considerably increases with pressure.
It increases even further if $\eta_1$ and $\eta_2$ become unequal, reaching several nm in Fig. \ref{figure4}b.
It is important to emphasise that $1/\tau$ is not the strain rate,
which can be estimated in our case as $(\rho(x\to\infty)^{-1} - \rho_0^{-1})/(\tau\rho_0^{-1})=(1-\gamma)/\tau$,
hence, being two orders of magnitude lower than $1/\tau$.

The intimate relation between the dilatational and interfacial viscosity coefficients determines the very ability of 2D water to flow.
This is the most non-trivial finding of this work. We confirm this finding by means of MD simulations in what follows.

\section*{Discussion}
\label{result}

\begin{figure*}
\includegraphics[width=\textwidth]{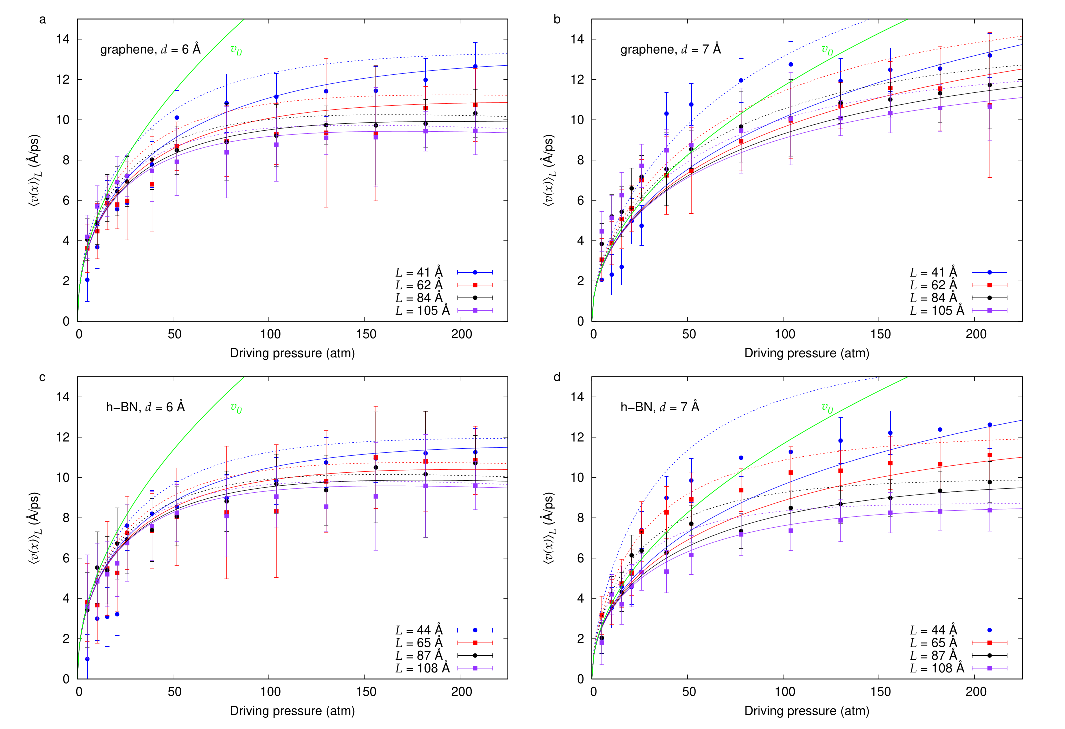}
 \caption{{\bf Hydrodynamic and molecular dynamic (MD) simulation results combined.}
 The solid curves represent the exact solution in terms of the Lambert functions (see ``Non-linear hydrodynamics in a 2D channel'' in  Methods),
 the dashed curves are given by Eq. (\ref{vLa}), and the thick green curve is the initial velocity $v_0$ given by Eq. (\ref{1st}) as an eye-guide.
 The averaged flow velocity is calculated from our MD data and fitted by the non-linear hydrodynamic model.
 {\bf a} Narrow graphene channel ($h = 2.66 \, \mathrm{\mathring{A}}$), MD data fitted with $\eta=6.5$ mPa$\cdot$s, $\eta_2=0.028$ mPa$\cdot$s.
 {\bf b} Wider graphene channel ($h = 3.66 \, \mathrm{\mathring{A}}$) requires higher viscosity values $\eta=10$ mPa$\cdot$s, $\eta_2=0.1$ mPa$\cdot$s to fit MD data.
 {\bf c} Narrow h-BN channel ($h = 2.66 \, \mathrm{\mathring{A}}$), MD data fitted with $\eta=3.9$ mPa$\cdot$s, $\eta_2=0.020$ mPa$\cdot$s, which are smaller than the values
 required to fit the data for graphene channel of the same height.
 {\bf d} Wider h-BN channel ($h = 3.66 \, \mathrm{\mathring{A}}$), MD data fitted with $\eta=35$ mPa$\cdot$s, $\eta_2=0.1$ mPa$\cdot$s.
 The remaining parameters are the same in all panels: $\rho_0=1\cdot 10^3$ kg$\cdot$m$^{-3}$, $c_0=1.5\cdot 10^3$ m$\cdot$s$^{-1}$,
 $H = 30\, \mathrm{\mathring{A}}$, which result in a Reynolds number $Re= \rho_0 v_0 h/\eta$ of the order of 0.001.
The colour curves approach the green one at larger $d$ indicating a gradual transition to the conventional regime
with the flow rate described by the Bernoulli relation (\ref{1st}). The error bars represent the standard deviation.
}
 \label{figure5}
\end{figure*}

We compare the velocity averaged over the length of the channel obtained from the hydrodynamics model with that obtained from molecular dynamics simulations.
The error bars of the simulated velocity $v(x)$, given by the standard deviation of the velocities of the molecules and intrinsic to the atomistic description,
are too large to allow us to analyse the velocity profile. Instead, we consider the velocity averaged over the length of the channel,
$\langle v(x) \rangle_L$, see Methods.  We plot the simulated and predicted $\langle v(x) \rangle_L$  as a function
of driving pressure for different channel lengths and materials the 2D channel can be fabricated from, see Fig. \ref{figure5}.
Note that we are not able to distinguish between the interfacial and dilatational effects within our non-equilibrium MD simulations. Instead, we {\it fit} the MD data by adjusting $\eta$ and $\eta_2$.
The ratio $\eta_2/\eta$ determines the sensitivity of $\langle v(x) \rangle_L$ to $L$.
If $\eta_2/\eta \sim 1$,
then the curves $\langle v(x) \rangle_L$ plotted for different channel length $L$ are indistinguishable at reasonable pressures.
Since our MD data suggests a certain dependence of $\langle v(x) \rangle_L$ on $L$ we use the ratio $\eta_2/\eta$
to fit the difference between the curves for the shortest and longest channels.
The absolute values of the viscosity coefficients are chosen to fit the dependence on driving pressure.

\begin{figure*}
\includegraphics[width=\textwidth]{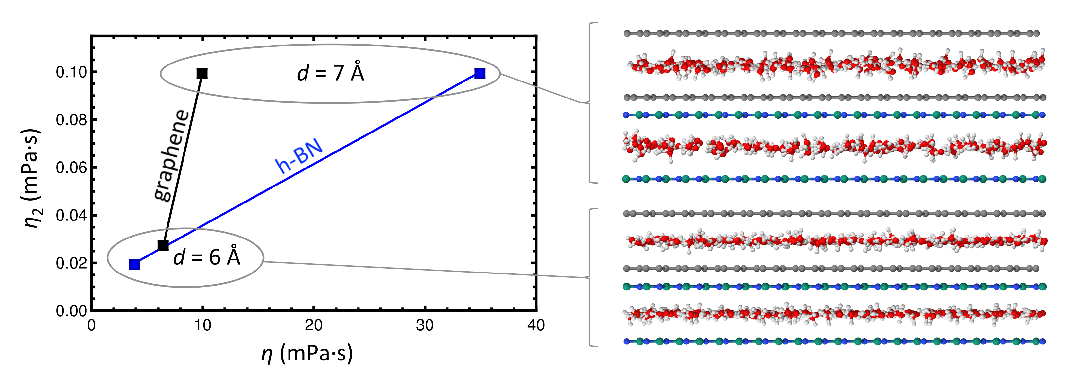}
 \caption{{\bf Viscosity parameter values for two-dimensional (2D) water confined by different materials.} Changing the channel's height and material strongly influences the interfacial and dilatational viscosity coefficients for 2D water.  The snapshots on the right show that while all oxygen atoms are nearly aligned in one plane in narrow channels (6 $\mathrm{\mathring{A}}$), they acquire an out-of-plane staggering pattern in wider channels (7 $\mathrm{\mathring{A}}$) leading to stronger interactions with the channel walls. The pattern is more pronounced in the hexagonal boron nitride (h-BN) channel, which is reflected in the higher interfacial viscosity coefficient. O, H, C, B and N atoms are represented in red, white, gray, green and blue, respectively.}
 \label{figure6}
\end{figure*}

It is instructive to have a simple algebraic expression for $\langle v(x) \rangle_L$ to understand its qualitative 
dependence on the length of the channel and on the driving pressure. Assuming high pressure and large length we obtain
\begin{equation}
 \langle v(x) \rangle_L  \approx  \frac{\frac{\eta_2}{\eta}v_0 \rho_0 c_0^2}{p_0 + \frac{\eta_2}{\eta}\rho_0 c_0^2}\left(1 + \frac{v_0 t_0}{2L}\right),
 \label{vLa}
\end{equation}
where $t_0=\tau/\gamma \equiv {\eta}/{\rho_0 c_0^2}$.
The dashed curves in Fig. \ref{figure5} show that the approximation works
reasonably well for longer channels in the region of relevant pressures.
Equation (\ref{vLa}) and Figure \ref{figure5} suggest that the $L$-dependence can be neglected if it is longer than $\sim 100\, \mathrm{\mathring{A}}$.
The dependence of $\langle v(x) \rangle_L$ on driving pressure is less trivial.
The high-pressure limit should be considered with great care because 
it may obviously result in a high flow velocity $v(x) > c_0$ making our explicit solution of Eq. (\ref{general}) inapplicable, see Methods.
Nevertheless, the pressure dependence is non-linear even if the velocity is relatively low.
Initially, $\langle v(x) \rangle_L $  increases with the driving pressure but eventually reaches the maximum and 
either drops back to zero in formally infinite channels or saturates in the finite channels.
We are not able to deal with sufficiently long channels within our MD framework, let alone the infinite ones,
hence, the flow velocity decrease is not visible in Fig. \ref{figure5}.
The saturation is however obvious.

The physical reason of such a non-linear behaviour is the unavoidable compression of 2D water upon its propagation
through the channel. The higher driving pressure results in stronger compression, more intensive energy dissipation, and 
higher resistivity against the water flow. 
The effect strongly depends on the channel height and material, see Fig. \ref{figure6}.
The general rule we find is as follows: The more ordered the water structure is, the lower viscosity coefficients are required
to fit the simulated water flow velocity profile.
However, the model involves two viscosity coefficients describing two different dissipation mechanisms. 

In wider channels ($d=$ 7 \AA), the 2D water structure experiences strong out-of-plane staggering, see the side-view snapshots in Fig. \ref{figure6}.
The staggered structure leads to stronger interactions between the water layer and the channel walls, increasing the interfacial viscosity coefficient $\eta_1$.
The effect turns out to be about 3 times stronger on h-BN than graphene, 
which agrees with the Green-Kubo estimations of the friction coefficients for bulk liquid water \cite{tocci2014friction}.
The dilatational viscosity coefficient $\eta_2$ remains the same ($\sim 0.1$ mPa$\cdot$s) in wider channels no matter which material is chosen.
It is interesting to note that the value $\sim 0.1$ mPa$\cdot$s equals the viscosity minimum deduced from the fundamental constants \cite{min-visc}.

In narrower channels ($d=$ 6 \AA), the water molecules are well aligned in the plane, Fig. \ref{figure6}.
The interfacial viscosity coefficient is therefore strongly reduced, up to one order of magnitude in h-BN channels.
The dilatational viscosity is also reduced by a factor of five, which suggests higher compressibility of truly 2D water. 
We therefore conclude that while the interfacial viscosity is associated with the out-of-plane staggering of 2D water layer,
the dilatational viscosity is mostly related to its in-plane structure.
Note that the in-plane structure is also determined by the out-of-plane interactions with the channel walls, so that
the two effects cannot be fully separated.

An increase of the channel height $d>7$\AA~ allows for even more disordered out-of-plane motion
developing a second water layer at $d\sim 1$ nm, see Supplementary Movie 3.
The resulting viscosity exhibits large oscillations as a function of $d$ originating
from commensurability between the channel height and the size of water molecules \cite{neek2016commensurability}.
We emphasize, however, that it is the dilatational viscosity $\eta_2$, rather than the interfacial one, that is responsible for the non-linear hydrodynamic features we found.

We note that h-BN is a polar crystal, in contrast to graphene.
Nitrogen accumulates excess electron charge leaving boron positively charged.
The resulting electrical polarisation creates an electric field which may influence the orientation of a water molecule nearby and potentially alter viscosity parameters.
The interaction of this polarisation with the water dipoles is one of the reasons for the different structures of water observed.
However, B and N atoms are stacked on top of each other in the AA$^\prime$ h-BN double-layers we consider here, except for the edge atoms, 
so that the in-plane electric field of each of these B-N pairs is compensated in the middle plane of the channel,
and there is only a smoothly varying electric field due to the edges.
Such is not the case for AA stacked h-BN double-layers, where the in-plane component of the electric field oscillates.
We performed similar MD simulations for AA stacked h-BN double-layers, Supplementary Figure 5, and observed expected
deviations from our hydrodynamic predictions because of the electric field fluctuations in the channel.
We have also considered hybrid channels made of h-BN and graphene, 
where the dipole fields are also not compensated in the middle plane of the channel, Supplementary Figure 6.
The deviations from our hydrodynamic predictions became much weaker but remain visible.
The dipole configurations are illustrated in Supplementary Figure 7.

Finally, we comment on possible experimental verification of our theory.
The 2D channels can be now fabricated through van der Waals assembly \cite{geim2013van},
with atomically flat sheets at the top and bottom \cite{2Dslit2016}.
The driving pressure of tens of atm can be created by an osmotic effect filling the left and right reservoirs
with pure water and strong sucrose solution, respectively \cite{andreeva2021two}.
Water flows have already been measured in graphene and BN nanocapillaries with spacing allowing for at least two water layers \cite{keerthi2021water},
and a similar method could be used to measure the viscosities of monolayer water.
Note that different materials will result in different structures of 2D water and different $\eta_{1,2}$.
This is an interesting opportunity to obtain various nanofluids out of the same water molecules
just by using alternate materials to fabricate the 2D channels.
It is worth emphasizing that the viscosity coefficients describing
2D water do not have the same meaning as for bulk \cite{jaeger2018bulk,tocci2014friction} and a-few-layer \cite{keerthi2021water} water.

A big open question is whether it is possible to reproduce the high selectivity and high permeability of natural aquaporins \cite{sui2001structural} 
by means of nanotubes \cite{holt2006fast,tunuguntla2017enhanced}.
MD simulations suggest the so-called single-file one-dimensional structure is formed
by water molecules in carbon nanotubes with a diameter of less than 1 nm \cite{hummer2001water,mukherjee2010single,su2011control} potentially facilitating permeability.
At the same time, the quantum mechanical charge fluctuation model \cite{kavokine2022fluctuation} suggests that the friction coefficient 
is strongly reduced in narrow carbon nanotubes,
as compared to graphite.
The natural channels are also short so that the major energy dissipation may occur right next to the channel entrance  \cite{gravelle2013optimizing}, similar to our theory. 
Further on, the natural channels combine  hydrophobic pores with specific hydrophilic sites. Such a structure is difficult to fabricate out of carbon nanotubes but
the hydrophilic/hydrophobic Janus-type 2D channels can probably be tailored out of two different 2D materials in a much simpler way. 
We, therefore, see 2D channels as potentially simpler structures to mimic biological functionalities of aquaporins.

\section*{Methods}

We apply a hydrodynamic description to the in-plane flow of 2D water, where
the basic hydrodynamic principles derived from the conservation of mass and momentum remain valid,
and take into account the nanofluidic effects by means of the viscosity parameters deduced from our MD simulations.

\subsection*{Non-linear hydrodynamics in a 2D channel}
\label{hydro}

The Navier-Stokes equation can be written as \cite{landau-hydrodynamics}
\begin{eqnarray}
\label{N-S}
 && \rho \left(\frac{\partial v_i}{\partial t} + 
  \sum\limits_k v_k \frac{\partial v_i}{\partial x_k}\right) =
 - \frac{\partial p}{\partial x_i} + 
 \sum\limits_k \frac{\partial \sigma_{ik}'}{\partial x_k}, 
\end{eqnarray}
where the viscous stress tensor is given by \cite{landau-hydrodynamics}
\begin{eqnarray}
\nonumber   
\sigma_{ik}' &= & \delta_{ik}\left[\eta_1\left(\frac{\partial v_i}{\partial x_k} + \frac{\partial v_k}{\partial x_i} - \frac{2}{3}
\sum\limits_l \frac{\partial v_l}{\partial x_l}\right)
+\eta_2 \sum\limits_l \frac{\partial v_l}{\partial x_l}\right]\\
&& + (1-\delta_{ik})\xi\left(\frac{\partial v_i}{\partial x_k} + \frac{\partial v_k}{\partial x_i}\right).
\label{sigma-ik}
\end{eqnarray}
Here, $t$ is the time, $i=\{x,y,z\}$ (as well as $k$ and $l$) are the coordinate indexes, and $\delta_{ik}$ is the Kronecker delta. We have introduced the first and second viscosity coefficients $\eta_{1,2}$ in the standard way so that 
the sum $\sigma_{xx}'+\sigma_{yy}'+\sigma_{zz}'$ does not depend on the first coefficient $\eta_1$ \cite{landau-hydrodynamics}. Besides, we single out the off-diagonal terms of $\sigma_{ik}'$ by introducing
the shear viscosity $\xi$ used to define the Navier partial slip boundary condition given by \cite{bocquet2010nanofluidics}
\begin{equation}
\left. \pm\xi\frac{\partial v_x}{\partial z}  \right|_{x=0,h}= \lambda v_x+o^2(v_x),
\end{equation}
where $\lambda$ is the friction coefficient.

The flow density and velocity must also obey the continuity equation given by \cite{landau-hydrodynamics}
\begin{equation}
\label{continuity}
 \frac{\partial \rho}{\partial t} + \sum\limits_k \frac{\partial v_k}{\partial x_k}=0.
\end{equation}

In what follows, the pressure gradient is 
applied along the $x$-direction, hence, $v_y=0$, $v_z=0$.
We are interested in a steady-state flow, hence, $\partial v_x/\partial t =0$ and $\partial \rho/\partial t =0$.
The Navier-Stokes equation then reads
\begin{eqnarray}
 \nonumber && \rho v_x \frac{\partial v_x}{\partial x} =
 - \frac{\partial p}{\partial x} + 
 \left(\frac{4}{3}\eta_1 + \eta_2\right) \frac{\partial^2 v_x}{\partial x^2} + 
 \xi\left(\frac{\partial^2 v_x}{\partial y^2} + \frac{\partial^2 v_x}{\partial z^2}\right). \\
 \label{N-S-2}
\end{eqnarray}
As 2D water maintains its in-plane structure we assume that $v_x$ does not depend on $y$ (the no-vorticity condition, $\mathrm{rot}\,\mathbf{v} =0$ within the water layer).
To eliminate the $z$-coordinate (hence, to approach the 2D limit) we average the shear viscosity term as
\begin{equation}
  \frac{1}{h}\int\limits_0^h dz \xi \frac{\partial^2 v_x}{\partial z^2} = -\frac{2\lambda}{h}v_x.
\end{equation}
Finally, we denote $v_x=v$, and Eqs. (\ref{N-S-2}, \ref{continuity}) then read
\begin{eqnarray}
\label{1} &&  \left(\frac{4}{3}\eta_1 + \eta_2 \right) \frac{\partial^2 v}{\partial x^2} - \rho v \frac{\partial v}{\partial x} - \frac{2\lambda}{h}v -  \frac{\partial p}{\partial x}
 =0,\\
\label{2} && \rho \frac{\partial v}{\partial x} + v \frac{\partial \rho}{\partial x} =0.
\end{eqnarray}
Note that Eq. (\ref{1}) does not depend on $\xi$ explicitly
because no intrinsic shear is assumed in structured 2D water.
The coefficients $\eta_{1,2}$ are retained in the 2D limit
but they are not intrinsic anymore; different from $\xi$, which reflects the interaction
between water layers in bulk water,  $\eta_{1,2}$ depend on the interaction between the water monolayer and the walls.
Thus, their values differ from those of shear and bulk viscosities in bulk water and in confined multi-layer water.

From Eq. (\ref{2}) we have $\rho(x)=\rho_0 v_0 /v(x)$.
Using ${\partial p}/{\partial x} = c_0^2 {\partial \rho}/{\partial x}$ we obtain
\begin{equation}
\frac{\partial p}{\partial x}= - \frac{c_0^2 v_0\rho_0}{v^2}\frac{\partial v}{\partial x},
\label{pressure-x}
\end{equation}
and Eq. (\ref{1}) then reads
\begin{equation}
 \label{general-friction}
 \left(\frac{4}{3}\eta_1 + \eta_2 \right)  \frac{\partial^2 v}{\partial x^2} + v_0\rho_0 \left(\frac{c_0^2}{v^2} - 1\right) \frac{\partial v}{\partial x}
 - \frac{2\lambda}{h}v=0.
\end{equation}
If the channel is short and the fluid is compressible, then the friction term can be neglected,
i. e. the last term in Eq. (\ref{general-friction}) is substantially smaller than the first two.
To validate this assumption for 2D water we introduce the critical channel length, $L_c$, and estimate 
the first and second velocity derivatives as ${\partial v}/{\partial x} \sim v/L_c$ and ${\partial^2 v}/{\partial x^2 \sim v/L_c^2}$,
respectively. Estimating $L_c$ by order of magnitude from Eq. (\ref{general-friction}) we set $v\sim v_0$, assume $\eta_1 \gg \eta_2$, and neglect the multipliers of the order of 1. The result reads
\begin{equation}
\label{Lc-eq}
  \frac{\eta_1}{L_c^2}  +\frac{\rho_0 v_0}{L_c}\left(\frac{c_0^2}{v_0^2}-1\right)-\frac{\lambda}{h}=0,
\end{equation}
which is a simple quadratic equation with respect to $L_c$.
The solution reads
\begin{equation}
 L_c=\frac{\rho_0 h\left(c_0^2-v_0^2\right)+\sqrt{4hv_0^2\eta_1\lambda + h^2 \rho_0^2\left(c_0^2-v_0^2\right)^2}}{2v_0\lambda}. 
 \label{Lc}
\end{equation}
If the actual channel length, $L$, is much larger than $L_c$, then the friction term dominates. In the opposite limit of short channels, $L \ll L_c$, the friction term can be neglected.
The critical length decreases when $v_0$ increases approaching $c_0$
so that $L_c>\sqrt{h\eta_1/\lambda}$ at $v_0<c_0$.
The hydrodynamic friction on graphene and h-BN is governed by classical mechanisms with negligible quantum corrections
\cite{kavokine2022fluctuation} resulting in a maximum
$\lambda^\mathrm{max}= 3\cdot 10^3$ N$\cdot$s/m$^3$.
Having in mind our channels with $h\approx 3$ \AA\,
and $\eta_1\sim 10$ mPa$\cdot$s we obtain
$\sqrt{h\eta_1/\lambda^\mathrm{max}}\sim 30$ nm
setting the lowest possible $L_c$ in the channels like ours.
Hence, considering 2D channels shorter than 100 \AA\, we can neglect the friction term in Eq. (\ref{general-friction})
and arrive at Eq. (\ref{general}) --- an intrinsically non-linear differential equation for compressible 2D water in short channels.

It is important that the non-linearity survives even in the limit ${c_0^2}/{v^2} \gg 1$ when the equation takes the form
\begin{equation}
\label{low-v}
 \eta \frac{\partial^2 v}{\partial x^2} + \frac{v_0\rho_0 c_0^2}{v^2}  \frac{\partial v}{\partial x}=0,
\end{equation}
where $\eta = 4\eta_1/3 + \eta_2$. 
Solution of Eq. (\ref{low-v}) can be explicitly written through the Lambert function $W_0(x)$ satisfying the following relation
$$
\frac{d}{dz} W_0(x) = \frac{1}{z}\frac{W_0(z)}{1 + W_0(z)}.
$$
Imposing the boundary conditions (\ref{1st}) and (\ref{2nd}) we obtain
\begin{equation}
\label{solution}
 v(x) = v_0 \gamma \left\{ 1+ W_0\left[ 
 - \frac{ \gamma-1}{\gamma}
 \exp\left( - \frac{ \gamma -1 + \frac{x}{v_0 \tau}}{\gamma}\right)\right]\right\},
\end{equation}
where $\gamma$ and $\tau$ are given by Eqs. (\ref{gamma}) and (\ref{tau}).

To compare our non-linear hydrodynamic model with our nanofluidic MD simulations we consider the averaged velocity
\begin{eqnarray}
\nonumber  && \langle v(x) \rangle_L = \frac{1}{L}\int\limits_0^L dx v(x)\\
 \nonumber & = &  \frac{v_0 \tau}{2L t_0^2}\left\{2Lt_0 + v_0 \tau^2 \left[ 
 \frac{t_0}{\tau} -1 - W_0\left(\frac{t_0-\tau}{\tau} 
 {\mathrm e}^{\frac{t_0-\tau}{\tau}-\frac{Lt_0}{v_0\tau^2}}\right)\right]  \right. \\
 && \left.  \times 
 \left[ 1 + \frac{t_0}{\tau} + W_0\left(\frac{t_0-\tau}{\tau} 
 {\mathrm e}^{\frac{t_0-\tau}{\tau}-\frac{Lt_0}{v_0\tau^2}}\right)\right]
 \right\},
 \label{vL}
\end{eqnarray}
where $t_0=\tau/\gamma \equiv {\eta}/{\rho_0 c_0^2}$.  

Note that in the formally supersonic limit, ${c_0^2}/{v^2} \ll 1$, we arrive at the linear differential equation given by
\begin{equation}
\label{high-v}
 \eta \frac{\partial^2 v}{\partial x^2} - v_0\rho_0 \frac{\partial v}{\partial x}=0,
\end{equation}
with the trivial solution 
\begin{equation}
\label{supersonic-solution}    
v(x) = v_0 +\frac{p_0}{v_0\rho_0}\frac{\eta}{\eta_2}\left(1- \mathrm{e}^{\frac{v_0\rho_0}{\eta}x} \right), \quad v_0\gg c_0,
\end{equation}
satisfying the  boundary conditions (\ref{1st}) and (\ref{2nd}).
As $v_0\rho_0/\eta>c_0\rho_0/\eta \sim 10^9$ m$^{-1}$, the velocity $v(x)$ drops exponentially within much less than 1 nm, 
and the flow should be described by Eq. (\ref{low-v}) again.
Hence, the model tends to be in the low-$v$ but intrinsically non-linear regime even though we start from an unrealistically high pressure making the flow formally supersonic.

\subsection*{Molecular dynamics simulations}
\label{MD}
We modelled the flow of monolayer water using classical molecular dynamics.
The system consists of a periodic 3D simulation box with cross-section 32$\times$30$\, \mathrm{\mathring{A}}^2$ and length between 150 and 230~\AA.
On the left, a mobile graphene piston of section 32$\times$30$\, \mathrm{\mathring{A}}^2$ is used to push the water through a graphene channel with length $L$. We have considered interlayer distances $d$ of 6 and 7~\AA, and channel lengths $L$=41, 62, 84, 105~\AA. The top and down graphene layers have Bernal stacking.
We performed similar simulations for a channel of AA$^\prime$-stacked BN, with length and cross-section adjusted for the difference in lattice parameters ($L$=44, 65, 87 and 108 \AA). 

The edges of the BN layers at $x=0$ were nitrogen-terminated, whereas the edges at $x=L$ were boron-terminated. The polarity of the BN contributes to the ordering of the water layer \cite{negi2022edge}.

The simulations were performed using the LAMMPS (Large-scale Atomic/Molecular Massively Parallel Simulator) code \cite{plimpton1995fast}.
The water molecules were modelled using the reparameterised simple point charge model (SPC/E) model \cite{berendsen1987missing,van1998systematic,mark2001structure}.
One of the considerations leading to the choice of this water model was the availability of both water-carbon and water-BN interaction parameters. 
We have previously tested four water models and two models for water-BN interactions \cite{negi2022edge}.
The relative energy of different water clusters is found to be within 12 meV/molecule of the DFT (Density-Functional Theory) values.
This is comparable to the error of the machine learning model recently reported (10 meV/molecule) \cite{kapil2022first}.
The shear and bulk viscosity values for SPC/E water are  0.67 and 1.56 mPa$\cdot$s, respectively \cite{jaeger2018bulk}.
The stacking of water layers was not considered, because we focused on monolayer ice \cite{ghorbanfekr2020insights}.

The water-carbon interaction was modelled by a Lennard-Jones potential between oxygen and carbon atoms, with parameters $\epsilon_{\rm OC}$= 0.114 kcal/mol and  $\sigma_{\rm OC}$= 3.28~\AA~\cite{joly2011capillary}.
The calculated water contact angle (WCA) for this parameter combination is 80$^\circ$,
see Supplementary Note 1 and Supplementary Figures 8--9.
We observed the square and rhombic phases for monolayer water confined by graphene,
consistent with the experimental observation of the square phase by electron microscopy \cite{algara2015square}.

The water-BN interaction parameters were adopted from a recent study \cite{wagemann2020wetting}, for which we obtained a contact angle of 73$^\circ$,
see Supplementary Note 1 and Supplementary Figures 10--11.
We have previously compared the phase diagram of water obtained with different BN-water potentials \cite{negi2022edge}.

Long-range Coulomb forces were computed using the particle-particle particle-mesh (PPPM) method.
Water molecules were kept at a constant temperature of 300 K using a Nos\'e-Hoover thermostat with a damping constant of 10~fs (100 timesteps).
We have neglected the streaming velocity in the temperature calculation, which we show to be a good approximation by carrying out further calculations with a PUT 
(Profile Unbiased Thermostat)  \cite{PhysRevLett.56.2172}, see see Supplementary Note 2 and Supplementary Figure 12).
The graphene was kept static except for the piston. The piston is not coupled to a thermostat when integrating its equation of motion.
A timestep of 0.1 fs was used.

The stationary flow velocity was calculated by averaging the velocity of the water molecules in the channel region
after the first molecules have reached the end of the channel and a stationary flow has been established.
Note that both in the beginning of the flow and at the end of the flow, when water is running out in the left reservoir,
there are transient regimes, where the equations for stationary flow presented here do not apply.
Notably, at the end of the simulation, when the piston becomes very close to the reservoir walls, 2D water is naturally formed in the reservoir as well.
However such transient regimes are not studied in the present work.
Stationary velocities have been obtained by averaging the velocity over a window of 5~ps at the start of the stationary regime.
The pressure in the left reservoir was estimated from the constant total force applied to the piston atoms,
while the pressure in the right reservoir is considered to be approximately zero in the beginning of the flow. 
We confirmed that there was no vorticity in the flow (Supplementary Figure 13).

We do not consider the channels higher than 7 $\mathrm{\mathring{A}}$ in the main text because water deviates from
a monolayer structure above  $\sim 8$ \AA\, \cite{gao2018phase}. 
The channels lower than 6 $\mathrm{\mathring{A}}$ are not considered also because water molecules do not enter such channels at moderate pressures.
The graphene/boron nitride planes were kept immobile, as in this study we intend to focus on the water dynamics. If the graphene/boron nitride planes would have been allowed to relax, the inter-layer distance, and, therefore, the water confinement potential, would not have been constant over the channel length.

The radial distribution function for 2D is defined as
\begin{equation}
 \mathrm{rdf}_\mathrm{2D}= \frac{1}{N}\sum_i\frac{\langle n(\mathbf{r}_i,r)\rangle}{\Omega r dr \rho_{\rm 2D}},
 \label{rdf}
\end{equation}
where $n(\mathbf{r}_i,r)$ is the number of oxygen atoms at distance $r$ from oxygen atom $i$, and $\rho_{\rm 2D}$  is the average 2D density, and $N$ is the total number of atoms, and $\Omega=2\pi$. The rdf was calculated for each snapshot of the molecules in the 2D region of the capillary ($L$=105 or 108~\AA), and averaged over the stationary flow time window. Since the system has no translation symmetry along the $x$ direction, in order to minimise the error at the edges, if the distance $d_i$ between atom $i$ and the edge was less than $r$, we used instead the edge-corrected expression with $\Omega=\pi\left(1+2\mathrm{arcsin}(d_i/r)\right)$.

Trajectory animations were created using Visual Molecular Dynamics (VMD) \cite{HUMP96}.

\bibliography{water.bib}

\section*{Data availability}

The authors declare that the data supporting the findings of this study are available within the paper and its supplementary information files
(Supplementary Figures 1 -- 13, Supplementary Notes 1 --2, and Supplementary Movies 1 -- 3).

\section*{Acknowledgements}

This research is supported by the Ministry of Education, Singapore, under its Research Centre of Excellence award to the Institute for Functional Intelligent Materials (I-FIM, project No. EDUNC-33-18-279-V12).
The computational work was supported by the Centre of Advanced 2D Materials, funded by the 
National Research Foundation, Prime Ministers Office, Singapore, under its Medium-Sized Centre Programme.

\section*{Author contributions}

M.T. conceived the project, proposed the hydrodynamic model, and wrote the first draft. A.C. implemented MD simulations and analyzed nanofluidic data.
A.H.C.N. supervised the project and discussed the results. All authors contributed to writing the final manuscript.

\section*{Competing interests}

All authors declare no competing interests.

\section*{Additional information}

{\bf Supplementary information} Supplementary Figures 1 -- 13, Supplementary Notes 1 --2, and Supplementary Movies 1 -- 3 are available.

{\bf Correspondence} and requests for materials should be addressed to Maxim Trushin (hydrodynamic) or Alexandra Carvalho (nanofluidic).

\end{document}